\documentstyle[12pt,epsfig]{article}
\newcommand{\be}{\begin{eqnarray}}
\newcommand{\ee}{\end{eqnarray}}
\newcommand{\half}{{\textstyle\frac{1}{2}}}

\newcommand{\Pslash}{P\hspace{-.5em}/\hspace{.15em}}
\newcommand{\pslash}{p\hspace{-.5em}/\hspace{.15em}}

\newcommand{\fourint}[1]{\int\!\frac{d^4 #1}{(2\pi)^4}}

\begin{document}
%
%
\rightline{UNITUE--THEP--11/97}
\rightline{TU--GK--5/97}
\rightline{hep-ph/9706551}
%
\vspace{.5cm}
\begin{center}
\begin{large}
{\bf Diquark confinement in an extended NJL model$^\dagger$} \\
\end{large}
\vspace{1cm}
{\bf G. Hellstern$^{\rm 1}$, R. Alkofer and H. Reinhardt}\\
\vspace{0.2cm}
{\em Institute for Theoretical Physics \\
T\"ubingen University \\
Auf der Morgenstelle 14 \\
D--72076 T\"ubingen, Germany} \\[.5cm]
\end{center}
\vspace{1cm}
\begin{abstract}
\noindent
In a Nambu--Jona-Lasinio model supplemented with an infrared cutoff
in addition to the ultraviolet cutoff
we study the issue whether diquarks are confined when the model 
is extended beyond the rainbow-ladder approximation. The gap equation,
obtained  in a truncation scheme motivated via  a nontrivial quark-gluon
vertex function, is solved to determine the 
constituent quark mass if  chiral symmetry is spontaneously 
broken. In a second step, the Bethe-Salpeter equations for
mesons and diquarks beyond ladder approximation are derived,
taking care to preserve Goldstone's theorem in the pion 
channel. 
While the obtained masses of  pseudoscalar and vector mesons
are only moderately shifted compared to the values in 
ladder approximation,
we observe that scalar 
diquarks disappear from the physical spectrum and 
therefore are confined. For axialvector diquarks we observe
indications, that the same mechanism may also work, but
the NJL model allows no conclusive answer in this channel. 
\end{abstract}
\vspace{.5cm}
\vfill
\rule{5cm}{.15mm}
\\
\noindent
{\footnotesize $^\dagger$ Supported by BMBF 
under contract 06TU888 and Graduiertenkolleg ``Hadronen und Kerne''
(DFG Mu705/3}). \\
{\footnotesize $^{\rm 1}$ E-mail:
Gerhard.Hellstern@uni-tuebingen.de} \\
\newpage
%
%
%
\section{Introduction}
Although QCD is believed  to be the theory of strong interactions
hadron properties have not yet been calculated from 
this underlying fundamental theory. 
Thus it proves helpful to develop 
effective theories including  the characterizing   
features of QCD as e.g. asymptotic freedom, confinement and chiral 
symmetry including its breaking pattern.
A model solely built on chiral symmetry and 
leaving aside the question of confinement and the transition 
to the perturbative region
is the Nambu--Jona-Lasinio (NJL) model \cite{NJL61}, 
formulated with quarks as elementary
degrees of freedom interacting locally. Modeling  the gluon sector  
to incorporate quark confinement is done  
in the Global Color Model (for a recent
review see \cite{Tan97}). Both approaches, due to their construction
principles, are able to describe the lowest lying
meson states quite reasonably. Baryons then emerge as 
solitons of the effective
meson theory \cite{Alk96,Fra91}. However, the modeled gluon 
interaction also gives rise
to quark-quark correlations, leading to a picture of baryons as 
diquark-quark bound states \cite{Cah89,Rei90}
(a hybrid model combining the soliton with the bound state picture 
has been developed in ref. \cite{Zuc97}). 

Although diquarks have many appealing 
phenomenological aspects (see e.g. \cite{Diq3}) and are
useful tools to parametrize unknown and/or complicated structures,
there remains the question whether diquarks are realized in nature. Diquarks 
as constituents in a baryon have to be necessarily in a color 
antitriplet state to build a colorless baryon when interacting  with a third
quark. On the other hand,
an antitriplet state is forbidden as an asymptotic state 
in a presumably confining theory 
like QCD. When the lowest lying diquark states (scalar and axialvector)
are evaluated in the NJL model in ladder  approximation
they are not confined. Also in the  
Global Color Model, which leads to quark confinement if
certain effective quark-quark interactions (which basically have 
to provide 
enough interaction strength in the infrared) are used, diquarks are predicted 
to be ``observable'' particles \cite{Cah87}. 

In ref. \cite{Ben96} it has been conjectured that this fact is due to the
commonly applied
rainbow-ladder approximation and diquark confinement can be obtained 
when the quark Dyson-Schwinger equation as well as the diquark
Bethe-Salpeter equation are considered beyond rainbow and ladder
approximation, respectively. While these studies have been 
performed in the Munczek--Nemirovsky model \cite{Mun83},
assuming an especially  simple form for  the gluon 
propagator \footnote{In all these studies it is assumed that the 
quark-quark interaction is given solely by the gluon propagator
despite the fact that this is not quite correct in a non-Abelian
gauge theory. E.g. ghosts invalidate this identification and an 
infrared diverging ghost propagator drastically alters this 
picture \cite{Sme97}.},
it has also been claimed that the observed feature of diquark confinement 
is generic and in particular independent of the gluon propagator.
Redoing the calculations of \cite{Ben96} with a realistic gluon propagator
(e.g. with  the ones reported in  ref. \cite{Rob94}) 
would be a very complicated 
task. We therefore want to test the conjecture of diquark confinement 
in a model, which assumes a quite different but also simplified
gluon dynamics: 
Whereas in the Munczek--Nemirovsky model the gluon 
propagator is a delta function in momentum space leading to a constant 
propagator in coordinate space, in the NJL model the gluon propagator 
is assumed to be a delta function in coordinate space, leading to
a constant propagator in momentum space.
When working with a nonrenormalizable theory, divergent 
integrals have to be regularized, leaving the cutoff finite.
We will use the proper-time method suggested in \cite{Ebe96}, i.e.
besides the usual UV cutoff also an infrared cutoff is introduced to  remove
unphysical quark thresholds from correlation functions and therefore 
mimicing quark confinement.

The paper is organized as follows: In the next section we solve 
the gap equation (quark Dyson-Schwinger equation) beyond rainbow
approximation; 
in section 3 we  show how the kernel of the meson and 
diquark Bethe-Salpter equation have to be consistently 
constructed to
preserve Goldstone's theorem in the pion channel. 
The solution of the  
Bethe-Salpeter equations for pions ($0^-$), vector mesons ($1^+$),
scalar ($0^+$) and axialvector diquarks ($1^-$) beyond 
ladder approximation are presented and discussed. 
After summarizing our results we will close with an outlook.  
In the appendices we discuss the elimination of thresholds in the
generalized proper-time scheme and $N_c-$counting beyond ladder 
approximation.

\section{Gap equation beyond rainbow approximation}
\label{secgap}
In Minkowski space\footnote{All calculations are done
in Minkowski space; a Wick rotation is performed only to evaluate the final
integrals. We furthermore assume exact flavor ${\rm SU}(2)$ 
symmetry and suppress
all explicit flavor indices.}
the Dyson-Schwinger equation, determining 
the dressed quark propagator is given by
\be
i S^{-1}(p)&=&\pslash -\Sigma(p) \nonumber\\*
\Sigma(p)&=& m_0 +
g^2 i\! \fourint{k} D_{\mu \nu}(p-k) \frac{\lambda^a}{2}
\gamma^\mu S(k) \Gamma_\nu(p,k)\frac{\lambda^a}{2}. 
\label{qsde}
\ee 
It connects the quark propagator $S(p)$ with the gluon  
propagator $D_{\mu \nu}(p)$ and the quark-gluon vertex function
$\Gamma_\nu(p,k)$.
The generators of color ${\rm SU}(3)$, the Gell-Mann matrices,  are denoted 
by $\lambda^a/2$. In the NJL model the gluon propagator is assumed  
to be 
\be
g^2 D_{\mu \nu}(p) \equiv G g_{\mu \nu},
\label{gluonprop}
\ee
leading to an effective, nonrenormalizable local quark-quark 
interaction governed by the coupling strength $G$. 
When furthermore the quark-gluon vertex function is substituted by its 
perturbative expression ($\gamma_\mu$) one ends up with 
the quark Dyson-Schwinger 
equation of the NJL model in mean field or rainbow approximation.
As suggested in ref. \cite{Ben96}, a minimal way to go beyond
rainbow approximation is to include the vertex correction 
to first order in $G$:
\be
\frac{\lambda^a}{2} \Gamma_\mu(k,p) &=& 
\frac{\lambda^a}{2} \gamma_\mu
+G i\! \fourint{l} D_{\rho \nu}(p-l)\gamma^\nu \frac{\lambda^b}{2} S(l+k-p) 
\gamma_\mu \frac{\lambda^a}{2}S(l) \gamma^\rho \frac{\lambda^b}{2} \\
\label{verrcorr1}
&=&
\frac{\lambda^a}{2} \left[\gamma_\mu
+ \frac{\sf -1}{\sf 6} G \gamma_\mu \Gamma^{L}((k-p)^2)\right]
\label{vercorr}.
\ee
The part of the vertex function $\Gamma^L((k-p)^2)$, describing the dressing 
of the perturbative quark-gluon vertex 
is given by
\be
\gamma_\mu \Gamma^{L}((k-p)^2)=i\! \fourint{l}
\gamma_\nu S(l+k-p) \gamma_\mu S(l) \gamma^\nu.
\label{verl}
\ee 
leading  to the quark Dyson-Schwinger equation
\be
\Sigma(p)= m_0 &+&
\frac{\sf 4}{\sf 3} G i\! \fourint{k}
\gamma_\mu S(k) \gamma^\mu \nonumber\\*
&+& \frac{\sf -2}{\sf 9} G^2 i\! \fourint{k}
\gamma_\mu S(k)\gamma^\mu \Gamma^L((k-p)^2).
\label{sigma}
\ee 
Throughout the paper we will denote factors, arising from the 
color structure by ``{\sf sans serif}'' letters. 
\begin{figure}[t]
\centerline{{
\epsfxsize 10.5cm
\epsfbox{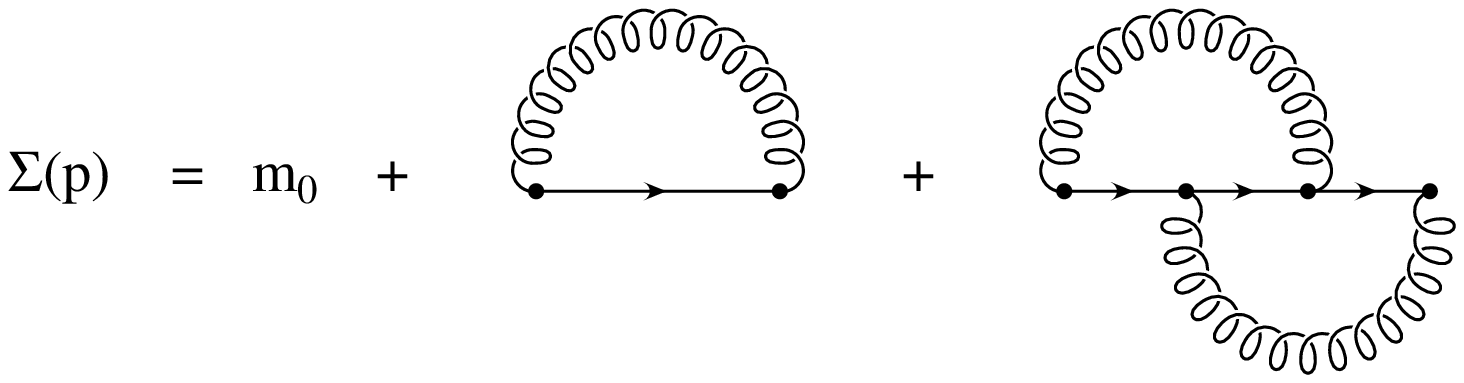}}}
\caption{{\it The quark Dyson-Schwinger equation  including the 
nontrivial  quark-gluon vertex function is shown.
Note, that due to the ansatz of the gluon propagator (eq. (\ref{gluonprop}))
the gluon lines are momentum independent and 
merely represent the color flow through the diagrams.} 
\label{dsefig}} 
\end{figure}
In fig. (\ref{dsefig}) the quark Dyson-Schwinger equation beyond rainbow
approximation is shown. Note, the gluon lines appearing in the
Feynman diagrams are given by eq. (\ref{gluonprop}) and are 
therefore momentum independent.

Here and in the following 
we regularize loop integrals in a generalized proper-time
scheme,
\be
\frac{1}{N^n}=\frac{1}{(n-1)!}\int_0^\infty \! ds s^{n-1}
\exp(-s N) 
\rightarrow \frac{1}{(n-1)!}
\int_{\frac{1}{\Lambda^2}}^{\frac{1}{\mu^2}}ds s^{n-1}\exp(-s N),
\label{propt}
\ee
where $N^n$ is the denominator of a Wick rotated 
loop integral after introducing the appropriate 
Feynman parameter integral and performing a corresponding momentum shift.
As has been shown in ref. \cite{Ebe96} any 
nonvanishing infrared cutoff $\mu$ (with $\mu \ll \Lambda$, where
$\Lambda$ is the ultraviolet
cutoff) removes unphysical quark thresholds from 
correlation functions and therefore simulates quark confinement in a crude
way, for details we refer to Appendix A.  
Note, however, although this prescription allows to investigate 
correlation functions at external momenta $P^2 > 4 M^2$, i.e. above
the ``pseudo''threshold, without opening the unphysical 
decay channel, this 
quark confinement mechanism is certainly too na\"\i ve to explain
physics far above the pseudothreshold. 
%
%
\noindent
\begin{table}[t]
\vspace{0.2cm}
\centering
\begin{tabular}{||c||c|c||c|c||}
\hline
& \multicolumn{2}{|c||}{\rm Rainbow, $O(G)$} 
& \multicolumn{2}{|c||}{\rm Beyond Rainbow, $O(G^2)$}\\ 
\hline
& $m_0 =0$& $m_0=0.017 $ &$m_0=0 $  &$m_0=0.017 $ \\ 
\hline
$\mu=0.1$ & 0.3873  &  0.400    & 0.4194    & 0.4295     \\
\hline
$\mu=0.2$ & 0.3873  &  0.400    & 0.4194    & 0.4297      \\ 
\hline
\hline
\end{tabular}
\\
\caption{\label{tabmass}{\it
The solution of the gap equation in rainbow $(O(G))$ and beyond
rainbow approximation $(O(G^2))$ is shown. We used parameters, which are 
fixed in the standard NJL model:  
The chosen set ($\Lambda=0.630\, {\rm GeV}, G=184.18\, {\rm GeV}^{-2}, 
m_0=0.017\, {\rm GeV}
$) lead to 
$M=0.4\, {\rm GeV} $ and to the pionic observables 
$m_\pi=0.140\, {\rm GeV} $ and $f_\pi=0.093\,{\rm GeV}$, 
when  the dominant amplitude of the pion (see eq. (\ref{pion}) is
considered) \cite{Wei93}. In the chiral limit the same parameters
are used, but $m_0=0$. 
All masses and scales are given in units of {\rm GeV}.}}
\end{table}
Due to the  gauge invariance of the proper-time 
regularization, we especially find
that the vertex correction is purely longitudinal (see eq. (\ref{vercorr})). 
We furthermore  observe that the 
vertex correction can be very well approximated by a simple 
dipole 
\be
\Gamma^L(Q^2) \sim \frac{a}{(Q^2+b)^2}, 
\label{dipol}
\ee 
with coefficients $a(M,\Lambda,\mu)$ and $b(M,\Lambda,\mu)$ 
obtained by a $\chi^2-$fit to the regularized result of eq. (\ref{verl})
which obviously depend on $\Lambda$, $\mu$ and $M$. 
Using the effective parametrization of the quark-gluon
vertex correction (\ref{dipol}) in eq. (\ref{sigma}) one finally 
ends up with the  
gap equation determining  the constituent quark mass
\footnote{$\Gamma(n,x)$ denotes the incomplete Gamma function arising
in the proper-time regularization scheme.}:
\be
M = m_0 &+& 
\frac{1}{16 \pi^2} \cdot \frac{\sf 4}{\sf 3} \cdot 4 M^2 G 
\left(\Gamma(-1,M^2/\Lambda^2)-\Gamma(-1,M^2/\mu^2)\right) \nonumber\\*
&+&
\frac{1}{16 \pi^2} \cdot \frac{\sf -2}{\sf 9} \cdot 4 M a G^2 
\left(\int_0^1 d x 
(1-x)\frac{1}{Y^2}(\exp(-Y^2/\Lambda^2)-\exp(-Y^2/\mu^2))
\right), \nonumber\\*
Y^2&=&(1-x)\cdot b +x M^2.
\label{gap}
\ee 
While the first line  of 
eq.(\ref{gap}) is the familiar
mean field result, the second line arises from the vertex correction.
Although the term induced by the vertex correction leads to 
a momentum dependent quark mass we assume that the momentum 
dependence at low energies is weak and it is therefore legitimate to
work with $\Sigma(p^2)\equiv M = const.$ in eq. (\ref{sigma}).
The gap equation (\ref{gap}) can be solved by taking 
the dependence of the dipole coefficients $a(M, \Lambda,\mu)$ 
and $b(M,\Lambda,\mu)$ on $M$ properly 
into account.
Our results 
are displayed in table (\ref{tabmass}).
It is seen, that the 
vertex correction is repulsive, leading to slightly larger
constituent quark masses. When doing  the calculation with different values 
of the infrared cutoff $\mu$, we find that the results are quite
insensitive on $\mu$, as long as there is enough phase space available 
i.e., the condition $\mu \ll \Lambda$ is fulfilled.

\section{Bethe-Salpeter equation for mesons and diquarks}
\label{secbse}
\begin{figure}[t]
\centerline{{
\epsfxsize 10.5cm
\epsfbox{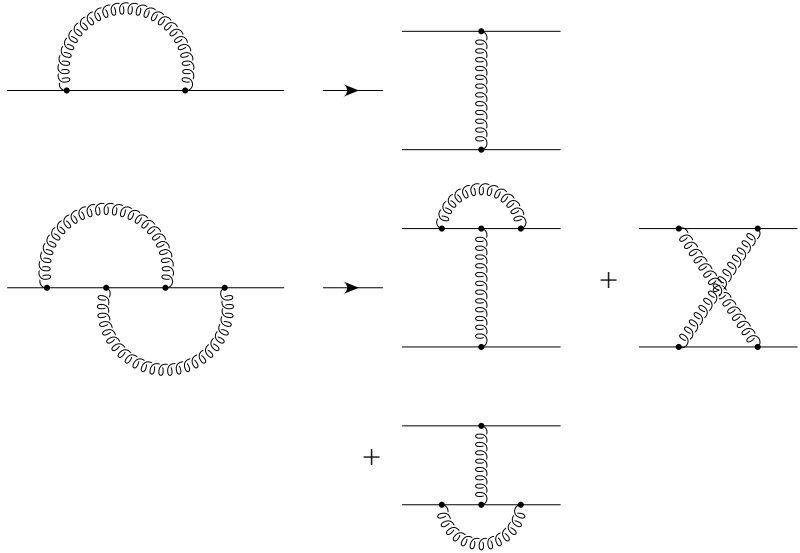}}}
\caption{{\it
In this figure we display how the kernel of the meson Bethe-Salpeter 
equation is obtained from the quark self-energy diagrams by cutting all
possible internal quark lines. See also fig.(1) of ref. \cite{Ben96}. 
} 
\label{diagram}} 
\end{figure}

The Bethe-Salpeter equation describing  mesons as quark-antiquark bound states 
is written as 
\be
\Phi_M(P,p)= \fourint{k} K_M(k,p;P) (S(k+\half P) \Phi_M(P,k) S(k-\half P)).
\label{mesonbse}
\ee
In ladder approximation the kernel $K_M(k,p;P)$ reduces itself to a momentum
independent contact interaction, reflecting the local quark-quark interaction
of the NJL model. It is well known, and can be proved quite easily,
that the ladder approximation preserves Goldstone's theorem: When  the 
constituent quark mass obtained from 
the gap equation  (eq. (\ref{gap}) without the vertex correction)
in the chiral limit  is used in eq. (\ref{mesonbse}), one
immediately finds 
massless bound states with  quantum numbers of  pions
(Goldstone bosons), reflecting spontaneously broken chiral symmetry.
Following  ref. \cite{Ben96} we construct the kernel
of the Bethe-Salpeter equation for mesons in a way which preserves
the Goldstone theorem in every order of the truncation scheme:
The Bethe-Salpeter kernel follows via the replacement \cite{Ben96}
\be
\gamma_\mu S(k) \gamma_\nu \rightarrow \gamma_\mu S(k+\half P) \Phi_M(k,P)
S(k-\half P)\gamma_\nu
\ee
from the quark self-energy $\Sigma(p)$. 
As it is displayed in fig. (\ref{diagram}), this corresponds to 
cut all internal quark lines in the quark self-energy diagrams, 
fig. (\ref{dsefig}), thereby
generating the Bethe-Salpeter kernel. Since the diagram in $O(G^2)$ 
contains three internal quark lines, the Bethe-Salpeter kernel
beyond ladder approximation consists of three parts.
This prescription leads to the  meson Bethe-Salpeter equation 
\be
\Phi_M(P,p)\!\!\!\!&=&\!\!\!\! \frac{\sf 4}{\sf 3} G_M i
\!\!\fourint{k} 
\gamma_\nu S(k+\half P) \Phi_M(P,k)  S(k-\half P)\gamma^\nu 
+\frac{\sf -2}{\sf 9} G_M^2 i \fourint{k} \fourint{l} 
\times \nonumber\\*
& & \left[ \right.  
\gamma_\nu S(k+\half P) \Phi_M(P,k)  S(k-\half P)\gamma_\rho 
S(l+k-p)\gamma^\nu S(l)\gamma^\rho \nonumber\\*
& & \mbox{} + \gamma_\nu S(k) \gamma_\rho S(l+k-p+\half P)\Phi_M(P,l+k-p)
S(l+k-p-\half P)\gamma^\nu S(l)\gamma^\rho  \nonumber\\*
& & \mbox{} + \gamma_\nu S(k) \gamma_\rho S(l+k-p)\gamma^\nu 
S(l+\half P)\Phi_M(P,l) S(l-\half P)\gamma^\rho \left. \right]. \nonumber\\*
\label{BSE}
\ee
By $G_M$ we collectively denote the coupling constants 
$G_M = \{G_\pi,G_{\rho, \omega}\}$ in the different meson channels.
While chiral symmetry forces the choice $G_\pi = G$ (see table
(\ref{tabmass})), the coupling 
constant in the vector meson channel 
can be different from that value. Nevertheless, we
use  $G_M=G_\pi=G_{\rho, \omega}=G$.  

Note that  the three terms  contributing to the
Bethe-Salpeter equation in $O(G_M^2)$ appear with an  equal 
color factor of ${-(N_c^2-1)/(4 N_c^2)=\sf -2/9}$, see
Appendix B. However, after
taking the relevant Dirac traces, the second term of order $G_M^2$
gets an additional minus sign, making this term repulsive 
compared to the first and third term which are attractive. 

In order to be consistent with the approximation employed in the 
gap equation, i.e. working with a  momentum independent
constituent quark mass,
we have to neglect the relative momentum of the Bethe-Salpeter amplitudes,
thereby using $\Phi(P,p)\equiv \Phi(P)$. For a pion
the appropriate Dirac structure of the amplitude is given by
\be
\Phi_\pi(P) = \gamma_5 \Psi_{\pi 1}(P) +\gamma_5 \Pslash \Psi_{\pi 2}(P).
\label{pion}
\ee
We therefore consider  not only the dominant Dirac structure 
$(\sim \gamma_5)$, 
but also the subdominant amplitude $(\sim \Pslash \gamma_5)$,
which is necessary to respect chiral symmetry and which induces
significant contributions to 
pionic observables (for a recent investigation
see ref. \cite{Bu97}). Physically, this is due to a
possible  $\pi-A_1$-mixing.
In case of vector mesons ($\rho,\omega$), the dominant transversal 
Dirac structure of the amplitude reads 
\be
\Phi^\mu_{\rho,\omega}(P) = (\gamma_\mu -\frac{P_\mu \Pslash}{P^2})
\Psi_{\rho, \omega}(P),\quad P_\mu \Phi^\mu_{\rho,\omega}(P) = 0.
\label{rho}
\ee  
Equivalently  one may use
\be
\tilde \Phi_{\rho,\omega}(P) = \epsilon^\lambda_\mu \gamma^\mu
\tilde \Psi_{\rho, \omega}(P),\quad  \quad\epsilon^\lambda_\mu P^\mu = 0,
\label{rho2}
\ee  
involving  the polarization vector 
$\epsilon^\lambda_\mu$ $(\lambda =0, \pm 1)$ being orthogonal to the total 
momentum of the vector meson.
A possible  subdominant amplitude would correspond to a mixing between vector
and tensor mesons, which are known to be absent in the NJL model.    
Therefore we have to restrict ourselves to the dominant Dirac structure
but choose it in a way which incorporates transversality from the 
very beginning.
After inserting the ans\"atze for the Bethe-Salpeter amplitudes
(eq. (\ref{pion}) and  eq. (\ref{rho}) or (\ref{rho2})) 
into eq. (\ref{BSE}), and taking
appropriate Dirac traces, the Bethe-Salpeter equation can be reduced 
to a system of two coupled algebraic equations for the pions 
and to one algebraic equation for the vector mesons, with the 
generic structure of the inverse propagators
of composite mesons
\be
[{\bf A}(P^2) - {\bf 1}]_{P^2=M_m^2}\Psi_m(P)=0,     
\label{disp}
\ee 
where $M_m = M_\pi \,\,{\rm or} \,\, M_{\rho,\omega}$. The meson masses
are given either by the zero of ${\it det} [{\bf A}(P^2) - {\bf 1}]$
(in the
pion channel, where ${\bf A}(P^2)$ is a $2 \times 2$ matrix) or 
simply 
by the zero of  eq. (\ref{disp}) (in the $\rho,\omega$ channel, where 
it is just one equation). 

In Appendix A we discuss the pion Bethe-Salpeter equation in ladder
approximation in order to demonstrate how the confinement 
mechanism via the infrared cutoff $\mu$ works. 

The two-loop integrals appearing in eq. (\ref{BSE}) are calculated 
as follows: First we perform one integration (e.g. over the momentum $l$) 
and find that the result (``longitudinal'' and ``transversal'' part)
can be  approximated by dipols, in analogy 
to eq. (\ref{dipol}); the second integration can then be performed  
using the dipols  as effective interaction between the quark 
and antiquark. 
The final 
expressions for the different ${\bf A}(P^2)$ are rather lenghty and 
will be explicitly given elsewhere \cite{Hel99}.

In order to calculate scalar and axialvector diquarks, the  
partners of the pseudoscalar and vector mesons, respectively, the
Bethe-Salpeter equation has to be modified in the following way.
The model independent expression of a diquark Bethe-Salpeter 
equation can be  written as 
\be
\Phi_D(P,p)= \fourint{k} K_D(k,p;P) (S(k+\half P) \Phi_D(P,k) S^T(k-\half P)),
\label{diquarkbse}
\ee
where ``T'' denotes the transposed of the Dirac matrix $S(p)$. 
As has been shown  in ref. \cite{Ben96}, each contribution 
to  the diquark kernel $K_D(k,p;P)$ can be obtained with the replacement
\be
S(p)\gamma_\mu \frac{\lambda^a}{2} \rightarrow [\gamma_\mu 
\frac{\lambda^a}{2}S(-k)]^T,
\label{trans}
\ee 
from the meson kernel $K_M(k,p;P)$ (eq. (\ref{BSE})) 
which leads to the diquark Bethe-Salpeter equation
\be
\label{DQBSE}
\Phi_D \!\!\!\!\! &(& \mbox{} \!\!\!\!\!\!\!
 P,p)=\frac{\sf -2}{\sf 3} G_D i
\!\!\fourint{k} 
\gamma_\nu S(k+\half P) \Phi_D(P,k)  S^T(-k+\half P)\gamma^{\nu T} 
\!\!+ G_D^2 i\!\! \fourint{k} \fourint{l}\!\! \times \nonumber\\   
&\left[ \right.&
\!\!\!\! 
\frac{\sf 1}{\sf 9}
\gamma_\nu S(k+\half P) \Phi_D(P,k)  S^T(-k+\half P)\gamma^T_\rho 
S^T(-l-k+p)\gamma^{\nu T} S^T(-l)\gamma^{\rho T} \nonumber\\
&\mbox{+}&
\!\!\!\! 
\frac{\sf -5}{\sf 9} 
\gamma_\nu S(k) \gamma_\rho S(l+k-p+\half P)\Phi_D(P,l+k-p)
S^T(-l-k+p+\half P)\gamma^{\nu T} S^T(-l)\gamma^{\rho T} \nonumber\\
&+& \mbox{}
\!\!\!\! 
\frac{\sf 1}{\sf 9} 
\gamma_\nu S(k) \gamma_\rho S(l+k-p)\gamma^\nu 
S(l+\half P)\Phi_D(P,l) S^T(-l+\half P)\gamma^{\rho T} \left. 
\right],
\ee
with $G_D = \{G_{0^+}, G_{1-}\}$. 
To match our result in the scalar diquark channel (when 
only the dominant Dirac structure is considered) with the ones  
obtained in a ladder NJL calculation \cite{Wei93}
we choose $G_{0^+}=2/3 G$.
To simplify comparison, in the axialvector channel we use
$G_{1^-}=2 G$, thereby obtaining in ladder approximation 
the same masses for 
axialvector diquarks and vector mesons.

When comparing the diquark Bethe-Salpeter equation 
with the one for mesons, the most striking 
difference are the different color factors 
in eq.(\ref{DQBSE}) arising from the projection onto 
color antitriplet states. While in  ladder approximation 
the Bethe-Salpeter equations only differ by a factor $2$, 
leading to a weaker binding of  diquarks when compared to mesons,
beyond ladder approximation these color factors are more significant.
The color prefactors of the
first and third term in $O(G_D^2)$ 
are given by $(N_c+1)/(2N_c)^2={\sf 1/9}$,
while for the second term 
one finds $(N_c+1)(1+N_c-N_c^2)/(2N_c)^2={\sf -5/9}$, see Appendix B.
Combined with factors arising from the Dirac structure of these
terms, 
the part 
$\sim {\sf -5/9}$ is repulsive, giving the possibility 
to avoid binding, thereby confining the diquarks.
It is amusing to note, that for $N_c=2$, the 
color factors for mesons and diquarks (which are then colorless ``baryons'')
are identical up to a minus sign, which is compensated by the different
Dirac structure. Unconfined mesons then automatically lead to unconfined 
diquarks, as expected. In the limit $N_c \rightarrow \infty$ only
the repulsive second term of eq. (\ref{DQBSE}), 
which is of $O(N_c)$, survives (the color prefactor in ladder 
approximation is given by $-(N_c+1)/2N_c=\sf{-2/3}$
, i.e. of $O(1)$), 
which is consistent  with Witten's conjecture \cite{Wit79}, stating 
that mesons are the relevant degrees of freedom for $N_c \rightarrow \infty$.
However, in QCD we have neither $N_c=2$ nor $N_c=\infty$, but 
$N_c=3$.

The ans\"atze of the diquark amplitudes 
in Dirac space are given by $\Phi_{0^+}(P) \simeq \Phi_\pi(P) C$
and $\Phi_{1^-}(P) \simeq \Phi_{\rho,\omega}(P)C$, 
respectively\footnote{$C$ denotes the charge 
conjugation matrix, obeying 
$C \gamma_\mu C^{-1}=-\gamma_\mu^T$. In the Dirac representation it is given
by $C=i\gamma^2 \gamma^0$.}.
When inserting these amplitudes into eq. (\ref{DQBSE}), the actual solution
of the Bethe-Salpeter equation is obtained in the same manner as for 
the mesons.

\section{Results and Discussions}
\label{secres}
\begin{figure}[t]
\centerline{{
\epsfxsize 10.5cm
\epsfbox{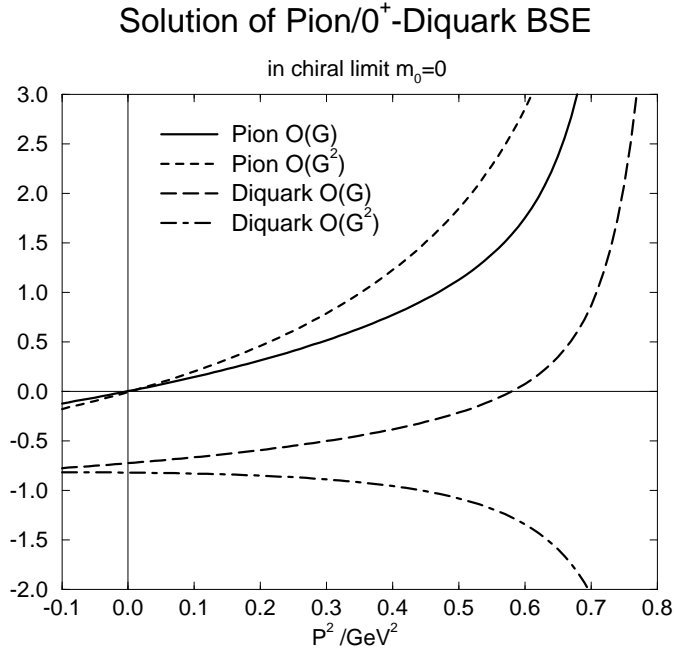}}}
\caption{{\it
The inverse of the propagators for the pion and the scalar diquark
are  shown in ladder $(O(G_{M,D}))$ and beyond ladder $(O(G_{M,D}^2))$ 
approximation in the chiral limit.
Its zeros  
correspond to the squared  bound state masses. 
Scalar diquarks are seen to be confined in $O(G_{0^+}^2)$.} 
\label{disp1}} 
\end{figure}
\begin{figure}[t]
\centerline{{
\epsfxsize 10.5cm
\epsfbox{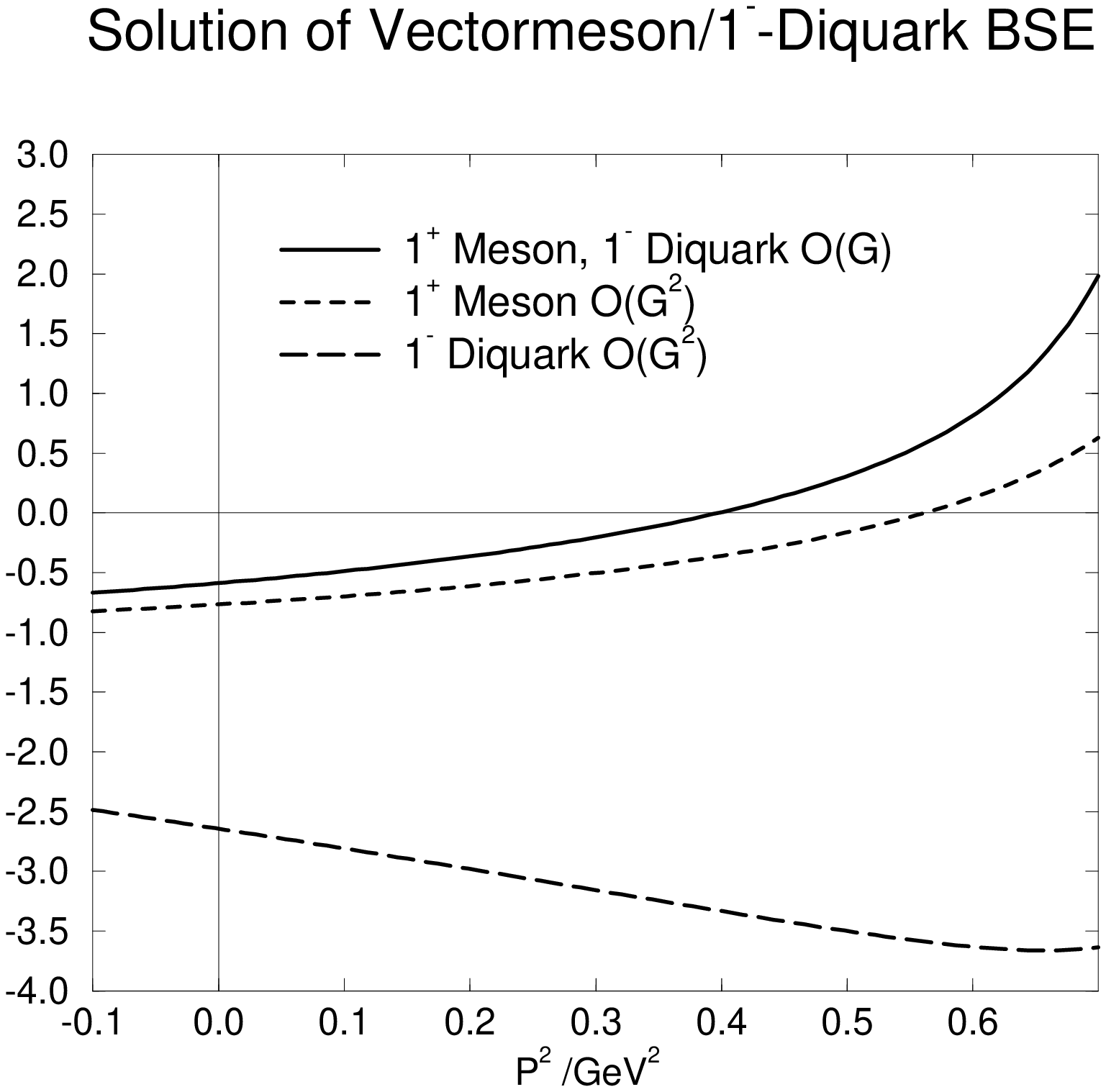}}}
\caption{{\it
The inverse of the propagators for the vector meson $(1^+)$
and the axialvector  diquark $(1^-)$
is shown in ladder $(O(G_{M,D}))$ and beyond ladder $(O(G_{M,D}^2))$ 
approximation.}
\label{disp2}} 
\end{figure}

%
%
\noindent
\begin{table}[t]
\vspace{0.2cm}
\centering
\begin{tabular}{||c||c|c||c|c||}
\hline
& \multicolumn{2}{|c||}{\rm Bound state mass in $O(G_{M,D})$} 
& \multicolumn{2}{|c||}{\rm Bound state mass in $O(G_{M,D}^2)$}\\ 
\hline
Quark Mass     & $M =0.3873 $& $M =0.40 $  &$ M =0.4194 $  &$M =0.4295 $ \\ 

\hline
Pion           & 0           &  0.207      & 0             & 0.182     \\
\hline
$0^+$ Diquark  & 0.762       &  0.786      & unbound       & unbound      \\ 
\hline
\hline
$\rho / \omega$& 0.612       & 0.630       & 0.737         & 0.749      \\
\hline
$1^-$ Diquark  & 0.612       & 0.630       & unbound       & unbound      \\ 
\hline
\hline
\end{tabular}
\\
\caption{\label{bsemass}{\it
In this table we show the bound state masses of the
mesons and diquarks in $O(G_{M,D})$ and $O(G_{M,D}^2)$. We use the 
parameters indicated in table {\rm(}\ref{tabmass}{\rm )} and values
of the coupling constants as explained in the text. The 
infrared cutoff is fixed to $\mu = 0.1 {\rm GeV}$. For
the diquarks ``unbound'' means, that the inverse of the propagators
has no zero below and slightly above
the pseudothreshold $P^2=4 M^2$, see figure {\rm(}\ref{disp1}{\rm)}, 
{\rm(}\ref{disp2}{\rm)}.}}
\end{table}
In this section we present and discuss our numerical results.
The bound state masses for the  mesons and diquarks 
in ladder ($O(G_{M,D})$) and beyond ladder ($O(G_{M,D}^2)$) 
approximation
are shown in table (\ref{bsemass}). 
As indicated in the previous section,
Goldstone's theorem is manifest for the pion; a massless 
bound state is found in the pseudoscalar channel when the 
gap equation and the Bethe-Salpeter equation are consistently  solved 
in the same order of $G=G_\pi$ (fig. (\ref{disp1})). 
When a small current mass 
is considered in the gap equation, i.e. chiral symmetry is
explicitly broken, a finite pion mass is obtained.
A comparison of the pion mass in order $G_\pi$ with the 
one obtained in order $G_{\pi}^2$,
entails slightly more attraction beyond ladder approximation.

For scalar diquarks we obtain the conjectured confinement 
effect: While to order $G_{0^+}$ bound state masses slightly 
below the pseudothresholds are  found, the large repulsive
term in the Bethe-Salpeter equation (\ref{DQBSE}) removes
the zeros from the inverse of the propagators
and therefore confines the scalar diquarks (see fig. (\ref{disp1})).
This behaviour is found in the chiral limit as well 
as in the calculation where a finite current quark mass is considered.
We observe furthermore, that the inclusion of the subdominant 
Bethe-Salpeter amplitude ($\sim \gamma_5 \Pslash C$) is 
necessary to obtain this feature. Working only with the
dominant amplitude leads  in  $O(G_{0^+}^2)$ 
to scalar diquark masses which, however, lies  above the pseudothreshold.   

While the pion (and also the scalar diquark) properties
are very much dictated by chiral symmetry constraints,
the situation is somewhat ambiguous in the vector meson
and axialvector diquark channel because there is no symmetry available
acting as a guiding principle for detecting the relevant structures.
In this respect, we find  that 
the results depend on details of the model, e.g.
the ansatz of the amplitude eqs. (\ref{rho},\ref{rho2})(one
might also think of using only $\gamma_\mu$ as the relevant Dirac structure 
of the amplitude), the values of the
coupling constants and even worse, the details of the regularization 
scheme (see the remark in Appendix A). 
Our choice of  $G_{\rho,\omega}=G$
and $G_{1^-}=2 G$ (leading to equal masses of vector mesons 
and axialvector diquarks in ladder approximation) and an ansatz of 
the amplitudes (\ref{rho}), which is ab initio transversal
to the total momentum allows at least a qualitatively 
comparison of vector meson and axialvector diquark properties
in  $O(G_{M,D}^2)$.

We then find that the vector meson mass in  $O(G_{\rho, \omega}^2)$ 
is about $100\, {\rm MeV}$ larger than the mass in $O(G_{\rho, \omega})$. 
For axialvector diquarks the inverse of the propagators has no zero 
below the pseudothreshold (see fig. \ref{disp2}), but the 
confining effect is not so clearly seen  as  in the scalar diquark 
channel, where the inverse of the propagator (as a function
of $P^2$) to $O(G_{0^+}^2)$ will obviously
never cross the $P^2-$ axis, see fig. (\ref{disp1}).
Here, however,
it may be possible, that far above the pseudothreshold a zero 
of the inverse of the propagator signals a bound diquark state. On the
other hand, as  mentioned 
in sect. (2), the model  should not be used in the region $P^2 \gg 4 M^2$.
Nevertheless, also in the axialvector diquark  channel
the repulsive term in the diquark Bethe-Salpeter equation 
has a significant effect, which may 
eliminate the axialvector diquark poles. Due to the
absence of tensor mesons (and therefore also tensor diquarks)
in the NJL model there is no a priori subdominant Bethe-Salpeter
amplitude for axialvector diquarks 
which could be included in the calculation and which may 
amplify the confining effect in analogy to the scalar 
diquark case. 

\section{Summary and Conclusions}
In this paper we investigated, in the 
context of a  NJL model with an infrared cutoff, the conjecture of ref.
\cite{Ben96} that diquarks may be confined in a truncation 
scheme, which is beyond rainbow-ladder approximation. 
For scalar diquarks we clearly found this feature,
due to the color structure a large
repulsive part of the Bethe-Salpeter kernel eliminates
diquark bound states, whereas in the pion channel, due to 
different color factors the same term is compensated by attractive
parts of the kernel. Vector meson masses are moderately shifted to higher
values relatively to their masses obtained in ladder approximation.
For axialvector diquarks it is  not possible 
to answer the question of confinement conclusively, but the 
observed behaviour also signals severe changes of the inverse diquark
propagator in this channel. To conclude, the proposed confinement
mechanism for diquarks also work in the NJL model in these
channels where the model is suitable: 
the pion and scalar diquark channel.

It will be interesting to see, if this mechanism of diquark
confinement  can be confirmed 
in a calculation including a realistic gluon propagator, thereby 
``interpolating'' between the NJL and the Munczek-Nemirovsky
model. 
 
Although the question how and whether diquarks are confined is 
an interesting que\-stion by itself, the outcome of such an investigation 
has to be included in a baryon calculation, where diquarks 
enter as effective constituents which are preferably confined.
Together with confined quarks such a description leads to 
baryons which have no quark-diquark thresholds any more \cite{Hel97}.
In such an effective diquark-quark model, besides the
diquark propagator also the diquark Bethe-Salpeter amplitude
enters, which can also be extracted from a diquark calculation
presented here or the one of ref. \cite{Ben96}.

\vspace{0.5cm}
\noindent
{\bf Acknowledgement}
We thank  L.\ v. Smekal for helpful discussions and critical 
reading of the manuscript. 
Useful comments by H.\ Weigel are gratefully acknowledged. 

\appendix
\section{Proper-Time Regularization with an Infrared Cutoff}
In order to explicitly demonstrate how the quark confinement mechanism
via an infrared cutoff \cite{Ebe96} works, we show here the solution 
of the pion Bethe-Salpeter equation in detail. For simplicity we will
restrict ourselves to the ladder approximation and consider only
the dominant Bethe-Salpeter amplitude in eq. (\ref{pion}).
The Bethe-Salpeter equation (\ref{BSE}) then reads
\be
\Psi_{\pi 1}(P) = 
\frac{\sf 4}{\sf 3} \frac{1}{4}  G_{\pi} i \fourint{k} \,{\rm tr}
\left[ 
\gamma_\nu S(k+\half P) \gamma_5 S(k-\half P)\gamma^\nu \gamma_5 \right]
\Psi_{\pi 1}(P).
\ee
Applying the standard techniques for the evaluation 
of one loop integrals leads, after
a Wick rotation,  
to the expression
\be
\Psi_{\pi 1}(P) &=& \frac{\sf 4}{\sf 3} 4 G_{\pi} 
\int_0^1 \!\! dx
\fourint{k_E}
\left[ 
\frac{1}{k_E^2 +M^2}+ \frac{(1-x)P^2}{(k_E^2+Y^2)^2}\right]
\Psi_{\pi 1}(P), \\
Y^2 &=& x(x-1)P^2+M^2.
\ee
Regularization is performed by using the prescription (\ref{propt})
to obtain
\be
\Psi_{\pi 1}(P) = \frac{\sf 4}{\sf 3}  4 G_{\pi} 
\int_0^1 \!\! dx \int_{\frac{1}{\Lambda^2}}^{\frac{1}{\mu^2}}\!\! ds
\fourint{k_E}
\left[
\exp(-s(k_E^2 +M^2))+ \right. \nonumber\\*
\left.
\left((1-x)P^2\right) s  \exp(-s(k_E^2+Y^2))
\right]
\Psi_{\pi 1}(P).
\label{bse2}
\ee
Note that $\mu^2=0$ is the usual NJL model expression in proper-time
regularization.  In eq. (\ref{bse2}) the integration over the 
loop momentum $k$ and over the proper-time $s$ can be done analytically and
the final expression (the integration over the Feynman parameter $x$
must be done numerically) for the pion Bethe-Salpeter equation is given 
by\footnote{When comparing this expression with the first line of
the gap equation (\ref{gap}) in the chiral limit one observes 
the manifestation of Goldstone's theorem.}
\be
\Psi_{\pi 1}(P) = \frac{1}{16 \pi^2} \frac{\sf 4}{\sf 3} 4 M^2  G_{\pi} 
\left[\left( \Gamma(-1,\frac{M^2}{\Lambda^2})-
\Gamma(-1,\frac{M^2}{\mu^2})\right)\right. \nonumber\\*
\left. + \frac{P^2}{2 M^2} \int_0^1\!\! dx 
\left( \Gamma(0,\frac{Y^2}{\Lambda^2})-\Gamma(0,\frac{Y^2}{\mu^2})\right)
\right]      
\Psi_{\pi 1}(P).
\label{bse3}
\ee
Here the $s-$integrals provided the expressions involving the incomplete
Gamma function, e.g., $\Gamma(0,Y^2/\Lambda^2)$ which becomes complex
for momenta $P^2$ larger than the threshold: The solutions of
$Y^2=0$ for a fixed $P^2$ are given by
\be
x_{1/2}=\frac{1}{2} \pm \sqrt{\frac{1}{4} -\frac{M^2}{P^2}},
\ee
i.e. below the threshold at $P^2=4M^2$ they are complex 
and therefore outside of 
the interval $[0,1]$, the integration region of the Feynman 
parameter $x$.
Above threshold, however, we find 
$Y^2 \le 0$, for  $0 \le x_1 \leq x\leq x_2 \le 1$, thereby
leading to an imaginary part of the incomplete Gamma function
\cite{AS65},
\be
{\rm Im}\,\Gamma(0,\frac{Y^2}{\Lambda^2})= \pm i \pi |x_2-x_1|
= \pm i \pi \sqrt{(P^2-4M^2)/P^2},
\ee
displaying the well-known square-root behaviour at the threshold.
Since the imaginary part is proportional to
$|x_2-x_1|$ but  independent of $\Lambda$, it is 
quite obvious that in the expression
\be
\int_0^1 \!\! dx \left(
\Gamma(0,\frac{Y^2}{\Lambda^2})-\Gamma(0,\frac{Y^2}{\mu^2})\right)
\label{diff}
\ee
which appears in eq. (\ref{bse3}) the imaginary part arising 
for $P^2 > 4 M^2$  is canceled 
if $\mu$ is kept nonzero.

Although the introduction of an infrared cutoff within the
proper-time regularization scheme seems to be 
a convenient way to avoid unphysical quark thresholds in correlation
functions, one must be very careful when applying this scheme in 
general. In ref. \cite{Bro96} it has been observed that 
$\Gamma(0,Y^2/\Lambda^2)$ has additional unphysical singularities in the
complex $P^2-$plane (which is quite
obvious, see \cite{AS65}), located on the imaginary $P^2-$axis
far beyond the applicability region of the model, $|P^2|<\Lambda^2$.
Since their location depends on $\Lambda^2$, they 
are not canceled by an infrared cutoff.
The expression  $\Gamma(0,Y^2/\mu^2)$
produces even more unphysical
singularities
in the correlation function, also located on the imaginary $P^2-$axis.
Although these singularities play no role if  only
the Bethe-Salpeter equation is solved, 
they certainly will affect e.g.
formfactor calculations, where a second external momentum can shift 
the singularities into the physically relevant region of the complex plane.
This indicates, that the proper-time regularization with an infrared cutoff
must be taken with great care.

Furthermore, we want to add the following remark: Although the 
proper-time regularization scheme is an 
unambiguous prescription when applied within a bosonization 
approach\cite{Ebe86} 
where the fermion determinant is regularized,
there are certain ambiguities if  the method is used on the level
of Feynman diagrams\footnote{The usual bosonization approach is 
deeply connected with the rainbow-ladder approximation and 
cannot be used in a truncation scheme like  the one considered in
this paper.}. 
These ambiguities are connected with 
various momentum shifts during the calculation. Since this 
problem does not arise in the gap equation (\ref{gap}), in the 
pion (and scalar diquark) channel one usually does the 
calculation in a way, which leads to an analytic
manifestation of Goldstone's theorem. 
However, in the vector meson (and axialvector diquark) channel,
chiral symmetry can not be used to resolve the ambiguities.
Therefore the results in these channels are not unambiguous.

\section{$N_c-$counting beyond ladder approximation}
In the meson
and diquark Bethe-Salpeter equations different color factors arise
which in turn lead to the differences in meson and diquark channels.
In color 
space, the meson Bethe-Salpeter (\ref{BSE}) equation can be denoted
by
\be
\varphi^M_{jk}= (\varphi^M_L)_{li} t^a_{ij} t^a_{kl} +
\left( (
 \varphi^M_1)_{li} t^a_{ij} t^b_{km} t^a_{mn} t^b_{nl} +
(\varphi^M_2)_{li} t^b_{im} t^a_{mj} t^b_{kn} t^a_{nl}+
(\varphi^M_3)_{li} t^b_{im} t^a_{mn} t^b_{nj} t^a_{kl}
\right)   
\ee
where the $(t^a)$ are  the generators of color ${\rm SU}(N_c)$.
The meson amplitude 
$\varphi^M_A,(A={L,1,2,3}$, where $L$ denotes the ladder part and
$1,2,3$ the three terms  beyond ladder approximation), 
describing a color singlet state, is given by the ansatz
$\varphi^M_{ij} =\tilde \varphi^M \delta_{ij}/\sqrt{N_c}$. Using this
ansatz leads to the 
following color factors:
\be
(L) &:& \frac{1}{N_c} \delta_{li} t^a_{ij} t^a_{kl} \delta_{jk} 
\quad\quad\quad= 
\frac{1}{2} \frac{N_c^2-1}{N_c}, \quad O(N_c)\\
(1) &:& 
\frac{1}{N_c} \delta_{li} t^a_{ij} t^b_{km} t^a_{mn} t^b_{nl} 
\delta_{jk}= - \frac{N_c^2-1}{4 N_c^2}, \quad O(1) \\
(2) &:& \frac{1}{N_c} \delta_{li} t^b_{im} t^a_{mj} t^b_{kn} t^a_{nl}
\delta_{jk}= - \frac{N_c^2-1}{4 N_c^2}, \quad O(1) \\
(3) &:& 
\frac{1}{N_c} \delta_{li} t^b_{im} t^a_{mn} t^b_{nj} t^a_{kl}
\delta_{jk} = - \frac{N_c^2-1}{4 N_c^2}, \quad O(1),
\ee
i.e. for  $N_c=3$ the factors appearing in eq.(\ref{BSE}) are 
obtained. In the spirit of an $1/N_c-$ex\-pan\-sion we observe that 
the term in ladder approximation would be of leading order $(O(N_c))$, 
while
the terms beyond ladder approximation are subleading, $(O(1))$.

The diquark Bethe-Salpeter equation (\ref{DQBSE})
in color space can be written 
as
\be
\varphi^D_{jk}= (\varphi^D_{L'})_{li} t^a_{ij} t^a_{lk} +
\left( 
(\varphi^D_{1'})_{li} t^a_{ij} t^b_{mk} t^a_{nm} t^b_{ln}+
(\varphi^D_{2'})_{li} t^b_{im} t^a_{mj} t^b_{nk} t^a_{ln}+
(\varphi^D_{3'})_{li} t^b_{im} t^a_{mn} t^b_{nj} t^a_{lk}
\right),   
\ee
where the prescription  (\ref{trans}) has been used to derive it
formally from the meson Bethe-Salpeter equation
(note, $(t^a_{ij})^T=t^a_{ji}$). The projection on color 
antitriplet diquark states is most conveniently done by using 
the ansatz
$\varphi^D_{ij} =\tilde \varphi^D \epsilon^a_{ij}/\sqrt{N_c-1}$ of the
diquark amplitude, involving the normalized Levi-Cevita tensor
$\epsilon^a_{ij}/\sqrt{N_c-1}$. Then the 
resulting color factors in the 
diquark channel are  given by
\be
(L') &:& 
\frac{1}{N_c-1} \epsilon^c_{li} 
t^a_{ij} t^a_{lk} 
\epsilon^d_{jk} \quad\quad\quad 
= -\frac{N_c+1}{2N_c}\,\delta^{cd},\qquad \qquad \qquad O(1)\\
(1') &:& 
\frac{1}{N_c-1} \epsilon^c_{li} 
 t^a_{ij} t^b_{mk} t^a_{nm} t^b_{ln}  
\epsilon^d_{jk}= \frac{N_c+1}{(2N_c)^2}\,\delta^{cd}, \qquad
\qquad \qquad O(1/N_c)\\
(2') &:& \frac{1}{N_c-1} \epsilon^c_{li} 
t^b_{im} t^a_{mj} t^b_{nk} t^a_{ln}
\epsilon^d_{jk}= \frac{N_c+1}{(2N_c)^2}(1+N_c-N_c^2)\,
\delta^{cd}, \quad O(N_c)\\
(3') &:& 
\frac{1}{N_c-1} \epsilon^c_{li} 
t^b_{im} t^a_{mn} t^b_{nj} t^a_{lk}
\epsilon^d_{jk}= \frac{N_c+1}{(2N_c)^2}\,\delta^{cd}, \qquad \qquad \qquad
O(1/N_c).
\ee
For the  case $N_c=3$ we obtain
the color factors appearing in eq.(\ref{DQBSE}).
As mentioned  in section(\ref{secbse}), the different terms 
in the diquark channel shows a rather interesting $N_c\rightarrow \infty$
behaviour. In leading order one would just 
keep the term $(2')$ which is of $O(N_c)$. However, the 
whole contribution of this term in the diquark 
Bethe-Salpeter equation is repulsive and thereby eliminates diquarks
totally. The attractive ladder part of the Bethe-Salpeter equation 
contributes  in this
scheme only in  subleading order, $O(1)$, and the attractive terms
beyond ladder approximation only in order $O(1/N_c)$.

\newpage
%
%

%
\end{document}